\documentclass[conference]{IEEEtran}
\IEEEoverridecommandlockouts
\usepackage[utf8]{inputenc}
\usepackage{graphicx}
\graphicspath{{figures/}}
\makeatletter
\def\input@path{{./sections/}}
\makeatother

\usepackage{cite}
\usepackage{amsmath,amssymb,amsfonts}
\usepackage{algorithmic}
\usepackage{graphicx}
\usepackage{textcomp}
\usepackage{xcolor}
\usepackage{balance}
\usepackage{url}
\usepackage{booktabs}
\usepackage{multicol}

\begin{document}
\bstctlcite{IEEEexample:BSTcontrol}

\title{The MIT Supercloud Workload Classification Challenge
\thanks{This material is based upon work supported by the Assistant Secretary of Defense for Research and Engineering under Air Force Contract No. FA8702-15-D-0001, and United States Air Force Research Laboratory Cooperative Agreement Number FA8750-19-2-1000. Any opinions, findings, conclusions or recommendations expressed in this material are those of the author(s) and do not necessarily reflect the views of the Assistant Secretary of Defense for Research and Engineering, or the United States Air Force. The U.S. Government is authorized to reproduce and distribute reprints for Government purposes notwithstanding any copyright notation herein.}}

\author{
        Benny J. Tang\IEEEauthorrefmark{1},
        Qiqi Chen\IEEEauthorrefmark{1},
        Matthew L. Weiss\IEEEauthorrefmark{2},
        Nathan Frey\IEEEauthorrefmark{2},
        Joseph McDonald\IEEEauthorrefmark{2},
        David Bestor\IEEEauthorrefmark{2},\\
        Charles Yee\IEEEauthorrefmark{2},
        William Arcand\IEEEauthorrefmark{2},
        Chansup Byun\IEEEauthorrefmark{2},
        Daniel Edelman\IEEEauthorrefmark{1},
        Matthew Hubbell\IEEEauthorrefmark{2},\\
        Michael Jones\IEEEauthorrefmark{2},
        Jeremy Kepner\IEEEauthorrefmark{2}, 
        Anna Klein\IEEEauthorrefmark{2},
        Adam Michaleas\IEEEauthorrefmark{2},
        Peter Michaleas\IEEEauthorrefmark{2},
        Lauren Milechin\IEEEauthorrefmark{1}, \\
        Julia Mullen\IEEEauthorrefmark{2},
        Andrew Prout\IEEEauthorrefmark{2},
        Albert Reuther\IEEEauthorrefmark{2},
        Antonio Rosa\IEEEauthorrefmark{2},
        Andrew Bowne\IEEEauthorrefmark{4}, \\
        Lindsey McEvoy\IEEEauthorrefmark{4}, 
        Baolin Li\IEEEauthorrefmark{3},
        Devesh Tiwari\IEEEauthorrefmark{3}, 
        Vijay Gadepally\IEEEauthorrefmark{2},
        Siddharth Samsi\IEEEauthorrefmark{2}\textsuperscript{\textsection} \\
    \IEEEauthorrefmark{1} MIT,
    \IEEEauthorrefmark{2} MIT Lincoln Laboratory,
    \IEEEauthorrefmark{3} Northeastern University,
    \IEEEauthorrefmark{4} US Air Force}

\maketitle

\noindent \copyright{ 2022 IEEE.} Personal use of this material is permitted. Permission from IEEE must be obtained for all other uses, in any current or future media, including reprinting/republishing this material for advertising or promotional purposes, creating new collective works, for resale or redistribution to servers or lists, or reuse of any copyrighted component of this work in other works.

\begingroup\renewcommand\thefootnote{\textsection}
\footnotetext{Corresponding author. Email : \url{sid@ll.mit.edu}}
\endgroup

\begin{abstract}
High-Performance Computing (HPC) centers and cloud providers support an increasingly diverse set of applications on heterogenous hardware. As Artificial Intelligence (AI) and Machine Learning (ML) workloads have become an increasingly larger share of the compute workloads, new approaches to optimized resource usage, allocation, and deployment of new AI frameworks are needed. By identifying compute workloads and their utilization characteristics, HPC systems may be able to better match available resources with the application demand. By leveraging datacenter instrumentation, it may be possible to develop AI-based approaches that can identify workloads and provide feedback to researchers and datacenter operators for improving operational efficiency. To enable this research, we released the MIT Supercloud Dataset, which provides detailed monitoring logs from the MIT Supercloud cluster. This dataset includes CPU and GPU usage by jobs, memory usage, and file system logs. In this paper, we present a workload classification challenge based on this dataset. We introduce a labelled dataset that can be used to develop new approaches to workload classification and present initial results based on existing approaches. The goal of this challenge is to foster algorithmic innovations in the analysis of compute workloads that can achieve higher accuracy than existing methods. Data and code will be made publicly available via the Datacenter Challenge website : \url{https://dcc.mit.edu}.
\end{abstract}

\section{Introduction}\label{sec:introduction}
High-Performance Computing (HPC) centers and cloud providers support a wide range of computational workloads, ranging from domain-specific scientific applications such as computational fluid dynamics, which represent applications written in high-performance programming languages such as C and Fortran, to Artificial Intelligence (AI)/Deep Learning (DL)/Machine Learning (ML) model training and inference. These diverse applications require different approaches to performance optimization as well as resource scheduling. HPC operators need new tools to understand the compute running on these systems and automated tools help identify mismatches between hardware capabilities and application requirements. To this end, workload classification helps to identify compute characteristics of unknown workloads and either present optimization pathways or alternatives to resource scheduling to minimize waste. This also enables potential approaches to job scheduling in a shared HPC cluster.  Wildani~\cite{wildani2015case} make a case for the development of workload classification techniques to better correlate workloads with system design requirements. While this work focuses on storage, understanding compute requirements has taken on a new urgency with the wide variety of hardware~\cite{reuther2021ai} available to HPC and cloud providers for the increasingly large AI/ML workloads which are typically written in high-level, interpreted languages such as Python or Julia, to name two. Thus, the goals of ensuring high performance, high availability, and efficient allocation of resources requires providers to  collect, integrate, fuse, and analyze the data of various system components such as storage, hardware, networking, applications, power, and other sensors from the cluster/datacenter. In particular, the analysis of compute utilization data from various compute workloads in a cluster has the potential to inform pathways to the optimization of cluster operations as well as user code. The recently released MIT Supercloud dataset~\cite{dcc-dataset} aims to provide such a curated dataset to enable these types of analyses. 

\subsection{Background and related work}\label{subsec:background}
Several prior efforts on workload classification and identification exist in the literature. For example, \cite{jiwoo2020} present results from workload classification on NERSC datasets. Approaches presented include feature selection followed by clustering algorithms, resulting in the categorization of jobs into three broad classes based on I/O characteristics. Energy consumption is a critical part of datacenter operations and \cite{sven2021} used power utilization signatures for workload classification. Yoo et. al.~\cite{yoo2016machine} used random forests to extract and characterize the patterns of unsuccessful job statuses based on known job characteristics. Part et. al.~\cite{park2017big} describe a scalable analytics framework applied to logs from the Oak Ridge Leadership Computing Facility. Klinkenberg et. al.~\cite{klinkenberg2017data} used sensor data from HPC systems to predict node failures. Finally, Banjongkan, et. al.~\cite{banjongkan2018multi} used a multi-label classification approach to predict job status based on HPC logs. Unfortunately, to the best of our knowledge, not all the data used for these analyses or implementations of methods are publicly available, which limits the wide-spread use of the proposed methods. To address this need, we are releasing a set of HPC monitoring logs from known workloads, referred to as \textit{the labelled dataset} herein, along with baseline implementations of our approaches to workload classification. While the initial dataset being released focuses primarily on AI/ML workloads, we intend to augment this dataset with other domain-specific scientific applications running in our cluster.  We also envision using subsets of MLPerf\cite{mlperf} as part of a larger labelled dataset.

Prior to the publication of the MIT Supercloud Dataset~\cite{dcc-dataset}, several HPC cluster and commercial cloud datasets have been made publicly available. These include the Parallel Workloads Archive (PWA)~\cite{pwa_web,pwa_paper}, the Grid Workloads Archive (GWA)~\cite{gwa_web}, and the Failure Traces Archive (FTA)~\cite{fta_web}.  Additional datasets include the Argonne Leadership Computing Facility~\cite{alcf}, Google Cluster-Usage Traces~\cite{google2010,google2011,google2019}, the Atlas Cluster Trace Repository~\cite{atlas2018}, the Philly-Traces~\cite{philly-traces} dataset from Microsoft, and the Blue Waters System Monitoring Data Set~\cite{bluewaters}. We refer the reader to the publications cited above for details on each dataset and availability. Along with these datasets, there is an increased interest in the analysis of cluster traces~\cite{prabhat2018lessons,baolin-hpca,alibaba,li-sc21} with a variety of goals such as developing better understanding of resource utilization and optimizing cluster operations to name a few. We hope that the MIT Supercloud Workload Classification Challenge (WCC) introduced in this paper will further address this need to understand HPC operations and foster new areas of research in this field. 

\subsection{Our Contribution}\label{subsec:contribution}
In this paper we present the results of machine learning classification algorithms on labelled HPC compute jobs as the baseline for the MIT Supercloud WCC. This is accomplished by utilizing a subset of the MIT Supercloud Dataset that consists of manually labelled AI training jobs.  The results presented here serve as a baseline for the MIT Supercloud WCC, which is the first public challenge as part of the larger MIT Datacenter Challenge (DCC).  For additional details, including data availability and updates on the MIT DCC, please see \url{https://dcc.mit.edu/}.  

The MIT Supercloud WCC, and the MIT DCC more broadly, is meant to foster the development of novel time series data pre-processing techniques and machine learning algorithms for datacenter operations.  We envision these innovations as tools which are generally applicable in any sensor environment where time series data are involved. In summary, the primary contributions of this paper are as follows:
\begin{itemize}
    \item Detailed presentation of the MIT Supercloud Dataset's labelled dataset
    \item Announcement of the MIT Supercloud Workload Classification Challenge
    \item Baseline implementations of the MIT Supercloud Workload Classification Challenge
\end{itemize}

\section{Dataset}\label{sec:dataset}
\subsection{The MIT Supercloud Dataset}\label{subsec:mit-sc-dataset}
The MIT Supercloud Dataset~\cite{dcc-dataset} was collected on the TX-Gaia  system, which is  a  heterogeneous cluster consisting of a set of GPU-accelerated nodes and another set of CPU-only nodes. The first partition has 224 nodes with two 20-core Intel Xeon Gold 6248 processors with a total 384GB of RAM and two NVIDIA Volta V100 GPUs with 32GB of RAM each. The dataset consists of time series of CPU and GPU utilization, memory utilization, GPU temperature, snapshots of compute node state, file I/O, as well as the scheduler log. All identifiable data has been removed or anonymized. Currently over 2.1 TB of data are available for download at \url{https://dcc.mit.edu/data}.  Further detailed descriptions of data collection, parsing, and anonymization are available in the paper~\cite{dcc-dataset}. 

\subsection{The Labelled Dataset}\label{subsec:labelled-dataset}
At present, among the over 2 TB of data in the MIT Superclould Dataset, approximately 2 GB consist of labelled workloads, with 3,430 unique jobs.  The labelled data was created by running and manually labelling commonly used deep neural networks in vision, Natural Language Processing (NLP) and Graph Neural Networks (GNN).  At present, there are ten deep neural network models in the labelled dataset as shown in Table \ref{tab:arch}.  However, many sub-architectures of the models in Table \ref{tab:arch} were collected as part of the labelled dataset.  This results in twenty six distinct classes of neural networks in the labelled dataset, as show in tables \ref{tab:vision-arch-1}, \ref{tab:vision-arch-2}, and \ref{tab:nlp-gnn-arch} in the Appendix.  Implementations of all models used to generate this data and the datasets are available for public download at \url{https://dcc.mit.edu/}.   The labelled dataset collection is ongoing and we expect to augment the existing data with additional AI and other HPC compute workloads.

\begin{table*}[t]
    \centering
    \caption{Architecture Totals for all Models}
    \begin{tabular}{lcclcclc}
    \toprule
        Vision Networks & Job Count & ~ & Language Models & Job Count & ~ & Graph Neural Networks & Job Count \\ 
        \cmidrule{1-2} \cmidrule{4-5} \cmidrule{7-8}
        VGG & 560 & ~ & Bert & 189 & ~ & DimeNet & 33 \\ 
        ResNet & 464 & ~ & DistillBert & 172 & ~ & SchNet & 39 \\ 
        Inception & 484 & ~ & ~ & ~ & ~ & PNA & 27 \\ 
        U-Net & 1431 & ~ & ~ & ~ & ~ & NNConv & 32 \\ 
        \bottomrule
    \end{tabular}
    \label{tab:arch}
\end{table*}

All data in the labelled dataset includes both CPU and GPU time series.  Due to the fact a single job may request multiple GPUs across multiple nodes, the number of distinct GPU time series is larger than 3,430.  For example, the datasets used in the experiments herein contain over 17,000 distinct GPU time series, although the labelling is repeated for a single job with multiple nodes and multiple GPUs.  
For completeness, modified versions of tables appearing in \cite{dcc-dataset} are included as Tables \ref{tab:cpu_clf_features} and \ref{tab:gpu_clf_features} here, which show the features in the CPU and GPU datasets relevant to classification.  Further details on the CPU and GPU time series datasets are available in \cite{dcc-dataset}.  

\begin{table*}[t]
    \centering
    \parbox{.48\linewidth}{
    \caption{CPU Time Series Features for Classification}
    \begin{tabular}{ll}
    \toprule
    \addlinespace[.1cm]
    \bf{Metric} & \bf{Description} \\
    \addlinespace[0.075cm]
    \midrule
    CPUFrequency & CPU clock frequency \\ 
    CPUTime & Time spent on compute by CPU \\ 
    CPUUtilization & CPU utilization by job \\ 
    RSS & Resident Memory Footprint Set Size \\ 
    VMSize & Virtual memory used by process \\ 
    Pages & Linux memory pages \\ 
    ReadMB,WriteMB & Amount of data read/written  \\ 
    \bottomrule
    \addlinespace[0.05cm]
    \end{tabular}
\label{tab:cpu_clf_features}
}
\hfill
\parbox{.5\linewidth}{
\caption{GPU Time Series Features for Classification}
\begin{tabular}{ll}
    \toprule
    \addlinespace[0.1cm]
    \bf{Metric} & \bf{Description} \\ 
    \addlinespace[0.075cm]
    \midrule
    utilization\_gpu\_pct & Percentage of GPU utilized \\
    utilization\_memory\_pct & Percentage of memory utilized \\
    memory\_free\_MiB & Available GPU memory \\ 
    memory\_used\_MiB & GPU memory in use \\ 
    temperature\_gpu & GPU temperature \\ 
    temperature\_memory & GPU Memory temperature \\ 
    power\_draw\_W & Power drawn \\ 
    \bottomrule
    \addlinespace[0.075cm]
\end{tabular}
\label{tab:gpu_clf_features}
}
\end{table*}

\section{Workload Classification Challenge}\label{sec:dcc}
The MIT Supercloud WCC is a supervised learning challenge with the goal of classifying labelled deep learning compute jobs running on the MIT Supercloud HPC.  As discussed above, the MIT Supercloud labelled dataset consists of 3,430 labelled jobs collected from an operational HPC system.  This presents many non-trivial challenges, some of which we discuss below.

\subsection{Challenge Datasets}\label{ssec:dcc:datasets}
As part of the WCC we are releasing seven datasets which are shown in Table \ref{tab:wcc_datasets}.  Note, the released datasets have been split into training and testing datasets with an 80/20 split ratio.  Each dataset contains approximately 60 seconds of GPU only data sampled from all trials in the labelled dataset that ran at least for (approximately) one minute.  The datasets were generated by sampling in three different ways: the first 60 seconds of each time series, the middle 60 seconds of each time series, and a 60 second sample drawn at random from the time series.  Each dataset is saved in the Numpy npz format and contains following the files: X\_train, y\_train, model\_train, X\_test, y\_test, model\_test.  Using the training sets as an example, X\_train is a three dimensional vector containing the time series data.  For example, in the 60-start-1 dataset, the dimensions are (14590, 540, 7) which correspond to trials, time series samples, and sensors respectively.  For all datasets, the seven sensors in the last dimension correspond to the seven sensors in Table \ref{tab:gpu_clf_features} and follow the same ordering as the table.  That is, element 0 is utilization\_gpu\_pct, element 1 is utilization\_memory\_pct, etc.  y\_train is a vector of integer class labels, with one label for each of the twenty six different architectures outlined in Tables \ref{tab:vision-arch-1}, \ref{tab:vision-arch-2}, and \ref{tab:nlp-gnn-arch} in the Appendix and model\_train contains the text names of the models corresponding to each numerical label in y\_train.

\begin{table}[!th]
    \centering
    \caption{Workload Classification Challenge Datasets}
    \begin{tabular}{cccccc}
    \toprule
        Dataset & Training Trials & Testing Trials & Samples & Sensors \\ 
        \cmidrule{1-5}
        60-start-1  & 14,590 & 3,648 & 540 & 7\\
        60-middle-1 & 14,213 & 3,554 & 540 & 7\\
        60-random-1 & 14,184 & 3,546 &  540 & 7\\
        60-random-2 & 14,183 & 3,546 &  540 & 7\\
        60-random-3 & 14,175 & 3,544 &  540 & 7\\
        60-random-4 & 14,193 & 3,549 &  540 & 7\\
        60-random-5 & 14,193 & 3,549 &  540 & 7\\
        \bottomrule
    \end{tabular}
    \label{tab:wcc_datasets}
\end{table}

\subsection{Formal Challenge Statement}\label{ssec:dcc:challenge}
Using the datasets discussed in Section \ref{ssec:dcc:datasets}, either individually or in combination, MIT Supercloud WCC submissions will be evaluated on classification accuracy, where the goal is to achieve an accuracy exceeding those presented in Sections \ref{sec:trad_ml} and \ref{sec:nn}.  

\subsection{Additional Considerations}\label{ssec:dcc:add-consid}
The proposed WCC is made difficult by the fact that not all time series have the same length.  Furthermore, given that the CPU and GPU time series are sampled at different rates, they will have different lengths for the same trial.  Solving the issue of aligning time series of varying lengths for machine learning is one of the primary problems this dataset presents.  This is made more complex by the fact the data was collected in a multi-sensor environment, as indicated by the number of features in Tables \ref{tab:cpu_clf_features} and \ref{tab:gpu_clf_features}.  Below are some further considerations posed by the labelled dataset and the MIT Supercloud WCC.

Given the number of samples in the labelled dataset, a neural network is likely to overfit. Can this be dealt with using regularization or resampling techniques?  Would traditional machine learning techniques be better suited for this problem?

Given the high data sampling frequency and long job run times the labelled dataset can be very large.  To this end, what kind of preprocessing and/or dimensionality reduction (if any) should be used?

Related to the previous point, determining feature importance may allow the exclusion of particular features without affecting classification accuracy.

Unlike traditional machine learning datasets, a single trial in the labelled dataset is not a vector in $\mathbb{R}^{n}$ but a matrix in $\mathbb{R}^{n \times m}$ matrix, $n$ being the number samples in the time series and $m$ the number of features.  While deep learning architectures such as Long Short-Term Memory Recurrent Neural Networks \cite{Hochreiter-lstm} address this issue, are there innovative data preprocessing techniques that map $\mathbb{R}^{m \times n} \to \mathbb{R}^{n}$ without sacrificing classification accuracy?

\section{Traditional Machine Learning - Baseline Models and Results}\label{sec:trad_ml}
In this section and Section \ref{sec:nn} we present baseline models and results from traditional machine learning and neural network architectures respectively.  The traditional machine learning models we experimented with were support vector machines (SVM), random forests (RF), and XGBoost, and a discussion of these appears in the subsections below.  All experiments were performed on the TX-Gaia system mention in Section \ref{subsec:mit-sc-dataset}.  Further details on the TX-Gaia system can be found in \cite{dcc-dataset}.  Additionally, the code used to generate the results below will be available at \url{https://dcc.mit.edu/}.

\subsection{Support Vector Machines and Random Forests}\label{subsec:svc_rf}
For both the SVM and RF architectures fourteen tests were performed.  This was the result of applying two different dimensionality reduction techniques to the seven datasets in Table \ref{tab:wcc_datasets}.  The first dimensionality reduction technique was principal component analysis.  As each trial in the datasets from Table \ref{tab:wcc_datasets} have 540 samples across 7 sensors, before performing PCA each trial was reshaped to have the dimensions 3,780.

As with any other type of sequential data, another challenge was the nature of time series encoding. For time series data, each column does not necessarily represent a stand alone feature, as any given point is extremely correlated with other samples temporally closer to it as well as points inside other periods. This requires feature engineering on the time series data for each layer as well as means to compare and contrast sequences across different sensors.

To address both of these challenges, the second dimensionality reduction technique was to compute the covariance matrix for each trial.  That is, given a single trial $\mathbf{M} \in \mathbb{R}^{540 \times 7}$, from either the training or testing sets, we computed the covariance matrix with respect to the seven sensors, $\mathbf{M}^{\top}\mathbf{M} \in \mathbb{R}^{7 \times 7}$.  As $\mathbf{M}^{\top}\mathbf{M}$ is symmetric, we further reduced the dimensions of each trial by taking the upper triangular portion of $\mathbf{M}^{\top}\mathbf{M}$.  These values were then stacked into a single row vector in $\mathbb{R}^{28}$.  This had the result of compressing the original training and testing datasets in $\mathbb{R}^{3}$ into datasets in $\mathbb{R}^{2}$.  For example, on the 60-start-1 training dataset, this technique maps $\mathbb{R}^{14590 \times 540 \times 7} \mapsto \mathbb{R}^{14590 \times 28}$, where 28 is the total number of unique variances/covariances that could be computed between seven sensors (the number of entries in the upper triangle of a matrix in $\mathbb{R}^{7\times 7}$). This significant reduction in dimensions helps reduce noise and thus allows easier detection of the real signal between sensor measurements and job types. It also allows us to directly assess the relevance of the relationship between sensors in the identification of job types.

For both SVM and RF, the best models were selected by performing a 10-fold grid search over a variety of hyperparameters. The hyperparameter specific to SVM was the regularization parameter C, with values of 0.1, 1.0, and 10.0.  For RF the hyperparameter was the number of trees, or estimators, having values of 50, 100, and 250. Both SVM and RF were implemented using Scikit-learn's \cite{scikit-learn} SVC and RandomForestClassifier classes respectively.  For the non-covariance datasets, the hyperparameter common to both models was the number of PCA dimensions, searching over the values 28, 64, 256, and 512.  For both SVM and RF, standardization was performed using Scikit-learn's StandardScaler class, with standardization being applied before either covariance or PCA dimensionality reduction.

The results for the SVM and RF experiments appear in Table \ref{tab:ml_results}, where the columns corresponding to the random datasets are labeled as R1, R2, etc.  With the exception of the 60-start-1 dataset, the best overall performing model was RF with covariance dimensionality reduction.  This is a significant result in that the time complexity for the covariance dataset, with a feature space in $\mathbb{R}^{28}$, was significantly less than the PCA datasets with larger feature spaces.  Another result worth pointing out is that, apart from the SVM model with PCA dimensionality reduction, the worst performance was achieved on the start dataset.  A possible explanation for this is the compute occurring at this time is not necessarily correlated uniquely with the specific neural network model being run.  For example, the data preprocessing and data loading phases occurring at this time may be generic across all models.  

\begin{table}[ht]
    \centering
    \caption{SVM and RF Test Accuracy (\%)}
    \begin{tabular}{lccccccc}
    \toprule
        Model & Start & Middle & R1 & R2 & R3 & R4 & R5 \\
        \midrule
        SVM PCA & 82.13 & 80.84 & 76.62 & 75.32 & 76.78 & 75.29 & 75.46\\
        SVM Cov. & 67.24 & 73.21 & 71.66 & 71.32 & 71.05 & 70.55 & 70.61\\
        RF PCA & \textbf{83.17} & 89.76 & 85.58 & 86.69 & 86.51 & 86.31 & 86.42\\
        RF Cov. & 81.80 & \textbf{93.02} & \textbf{90.05} & \textbf{90.64} & \textbf{90.01} & \textbf{90.73} & \textbf{90.90}\\
    \bottomrule
    \end{tabular}
    \label{tab:ml_results}
\end{table}
\subsection{Regularizing Gradient-Boosting Classifier}\label{subsec:grad_boost}

In addition to the above models, we also performed tests using the XGBoost algorithm, a type of sparsity-aware tree-boosting classifier. This recent ensemble method has gained great popularity in a variety of machine learning applications such as credit scoring, bioactive molecule prediction, and sentiment analysis \cite{xgboostcomp}, and its competitive performance has repeatedly placed it among the top contenders at Kaggle competitions \cite{chen-xgboost}. We evaluated its performance on the 60-random-1 dataset, applying standardization and covariance dimensionality reduction as in Section \ref{subsec:svc_rf}. The grid search consisted of 5-fold cross validation on the following hyperparameters.  The first was $\gamma$, which set the minimum loss reduction required to make a further partition on the leaf node of a tree.  Additionally, we performed a grid search over both $\alpha$ and $\lambda$ hyperparameters, which correspond to $\ell_{1}$ and $\ell_{2}$ regularization terms on the weights respectively.

Our XGBoost model achieved a test set accuracy of 88.47\% after 40 boosting rounds.  Further experiments determined that model performance plateaus after around 40 boosting rounds and the model is overfitting as the training set error is very close to zero.  Additionally, we were able to identify important covariances that have higher predicative power by comparing the feature importance value for each covariance. Feature importance indicates, on average, how frequently each attribute split point improves the accuracy metric. From this analysis we identified the top three sensor variances/covariances in terms of feature importance: 
\begin{itemize}
    \item Covariance between GPU \% Utilization and CPU \% Utilization
    \item Variance of GPU \% Utilization
    \item Variance of Power Draw
\end{itemize}

This result could lead to further investigations on how these GPU sensor measurements interact differently for each job type. For example, one could inspect the relative efficiency of the GPU in converting power to utilization for different job types by the corresponding magnitudes of measurements from the utilization GPU and power draw sensors, and contrast across different job types. This would give further insight on job efficiency on a more granular level.

\section{Recurrent Neural Networks - Baseline Models and Results}\label{sec:nn}
Recurrent neural networks (RNNs) have shown their efficacy in a variety of NLP tasks \cite{rnn-1} \cite{rnn-2} \cite{rnn-3}.
Similar to NLP tasks, our data consists of time series that are sequentially and temporally ordered, which traditional multi-layer perceptrons cannot properly capture.
Recurrent neural networks address this issue by sequentially processing sequences one step at a time, passing the output of one time step as the input of the next time step, making them suitable for our task at hand.
However, unlike traditional machine learning models, neural networks often require longer training times.  As a result, instead of training on all seven datasets in Table \ref{tab:wcc_datasets}, we trained on the 60-start-1, 60-middle-1, and 60-random-1 datasets.
Prior to training, each dataset was standardized using Scikit Learn's StandardScaler class as in Section \ref{sec:trad_ml}.
No other feature engineering or preprocessing was applied to the datasets.
\begin{table*}[t]
    \centering
    \caption{RNN Test Accuracy (\%)}
    \begin{tabular}{lccc}
    \toprule
        Model & Start Dataset & Middle Dataset & Random Dataset \\
        \midrule
        LSTM (h=128) & 82.57 & \textbf{92.09} & \textbf{90.81} \\
        LSTM (h=128, 2-layer) & 80.51 & 91.90 & 90.52 \\
        CNN-LSTM (h=128) & \textbf{82.65} & 89.90 & 90.55 \\
        CNN-LSTM (h=256) & 67.60 & 89.36 & 88.61 \\
        CNN-LSTM (h=512) & 64.45 & 65.67 & 73.80 \\
        CNN-LSTM (h=512, small kernel) & 66.26 & 71.47 & 75.21 \\
    \bottomrule
    \end{tabular}
    \label{tab:nnresults-combined}
\end{table*}
\subsection{Long Short-Term Memory Networks}\label{subsec:lstm}
A special type of RNN, Long Short-Term Memory networks (LSTM) are explicitly designed to address the long-term dependency problem in traditional RNNs \cite{Hochreiter-lstm}.
To establish a baseline comparison to classical machine learning approaches, we picked a fairly straightforward standard bidirectional LSTM architecture.
A bidirectional LSTM trains two LSTMs on the input sequence, one from start-to-end, and the other in reverse.
This allows the LSTM to learn additional long-term context of the input sequence from both past and future information, rather than just past information as in a unidirectional LSTM.

The input sequence is fed into a bidirectional LSTM with a hidden layer size of 128 with all 7 sensors as the feature vector.
Then, the output of both LSTMs were concatenated and passed through a fully-connected layer projecting down to a feature size equal to the length of the sequence.
This was then passed through a dropout layer with p=0.5 to reduce overfitting, a leaky rectified linear unit (leaky ReLU) \cite{maas2013rectifier} activation function as a non-linearity, and finally fed into another fully-connected layer that outputs a vector of size equal to the number of classes we are predicting.
Lastly, we apply a log-softmax transform on the output vector to get the log-probabilities of our classes.
For our loss function, we take the negative log-likelihood loss of the log-probability vector with respect to the correct classes.

During training, a cyclical learning rate scheduler was used with cosine annealing as it has been shown to drastically improve convergence and reduce training time \cite{lsmith-cyclic}. Each model was trained for 1000 epochs, early stopping if the validation accuracy did not improve over 100 epochs.
For consistency and simplicity, we report the best validation accuracy in our results rather than using a more nuanced, "best gut sense" manner of choosing the best epoch model that minimizes overfitting.  This process was then repeated for a stacked 2-layer bidirectional LSTM, with a dropout layer with p=0.5 in between, and all other hyperparameters held equal for a total of two LSTM models.

\subsection{Applying Convolutions to LSTMs}\label{subsec:cnnlstm}
As the seven GPU features in our dataset are correlated with one another, and our intuition informs us that we are likely to see certain patterns of GPU activity during workloads, we also introduce elements from the domain of computer vision and image recognition.
Convolutional networks (CNNs) \cite{lecun-lenet} "slide" filters (called convolutional kernels) across the dataset that allow it to learn intermediate features - in image processing tasks these might include learning to recognize facial features or parts of a vehicle in image recognition tasks.

We feed the input sequence into two 1-dimensional convolutional layers sandwiching a max pooling layer to reduce the dimensionality of the feature maps.
This output is then fed into the same bidirectional LSTM architecture from Section \ref{subsec:lstm} and trained in the same manner.
Because the training time of RNNs increases with the length of the input sequence, this had the side benefit of speeding up training time by almost 8 times.

We repeated this process with two more models with hidden sizes of 256 and 512, and a CNN-LSTM with a hidden size of 512, but with a smaller kernel and step size (and thus a longer sequence output length to be fed into the LSTM) for a total of four CNN-LSTM models.

The results of our RNN models are reported in Table \ref{tab:nnresults-combined}.  A comparison with Table \ref{tab:ml_results} shows that the best performing RNN models had marginally smaller accuracy values compared to the best models in Table \ref{tab:ml_results} on the 60-start-1 and 60-middle-1 datasets and performed slilghtly better on the 60-random-1 dataset.  Although the RNN models posses more degrees of freedom, and potentially can represent a larger class of functions, these more complex models may be overfitting to the training data and thus scoring lower in test.  Additionally, it is likely a more expansive hyperparameter sweep and augmented dataset would result in the RNN models outperforming the SVM and RF models. Nevertheless, our set of results represents a promising start in using modern neural approaches to tackle our challenge task of classifying workloads.

As with the SVM and RF results discussed above, the lowest RNN accruacy, for all models, was acheived on the 60-start-1 dataset.  This further supports our explanation of this same result in the context of the SVM and RF models from Section \ref{subsec:svc_rf}.

\section{Summary and Future Work}\label{sec:summary}
The results in Table \ref{tab:ml_results} are impressive given that SVM and RF are standard machine learning algorithms and certainly not considered state-of-the-art.  In fact, among all the baseline models evaluated herein, in general, the best accuracy was achieved with RF.  However, while this established a baseline, given the complexity of neural network architectures, it is quite possible greater diversity of data, modifications to the LSTM architectures, a more extensive hyperparameter search, or other neural network architectures may show improvement over this baseline.

Our results with XGBoost demonstrated that appropriately-engineered, preprocessed features in combination with a well-proven classical machine learning technique are competitive with state-of-the-art neural network architectures.  As with the LSTM architectures, the performance of XGBoost may be improved with more diverse data and a more expansive hyperparameter search.

Although deep neural networks have reduced the need for finely-crafted, low-level feature engineering, appropriate feature selection and high-level meta-structure features are still relevant and a worthy pursuit in our opinion.
As a result, we believe that the performance of our RNNs can be improved with the appropriate pre-processing and feature selection in our dataset.
Furthermore, given that our results suggest that our models might have overfit to the training data, more could be done to reduce overfitting, such as applying regularization or additional dropout layers.

Additionally, other more recent and novel neural network architectures may prove to be more accurate on our challenge dataset.
In particular, we believe that the ConvLSTM architecture is promising in its ability to capture convolutional features in both the input-to-state and state-to-state domains having shown efficacy in weather forecasting problems \cite{shi-convlstm}.

Lastly, we are excited by the prospect of training models on the entire dataset of workloads from start-to-finish.
While such models will take significantly longer to train than any of the ones we have used in this paper, we believe that the ability for them to learn the structures and patterns of a full workload will help in classifying snapshots of data from live workloads running in-progress, which represents a viable use case for these types of models to be deployed.  

\section*{Acknowledgments}
The authors acknowledge the MIT Lincoln Laboratory Supercomputing Center (LLSC) for providing HPC resources that have contributed to the research results reported in this paper. The authors wish to acknowledge the following individuals for their contributions and support: Bob Bond, Nathan Frey, Hayden Jananthan, Tucker Hamilton, Jeff Gottschalk, Tim Kraska, Mike Kanaan, CK Prothmann, Charles Leiserson, Dave Martinez, John Radovan, Steve Rejto, Daniela Rus, Marc Zissman.

\bibliographystyle{IEEEtran} 
\bibliography{IEEEabrv, references}
\section{Appendix}\label{sec:appendix}
\begin{table}[h]
    \centering
    \caption{VGG and Inception architectures used for vision models}
    \begin{tabular}{lcclc}
    \toprule
        VGG Models & Job Count & ~ & Inception Models & Job Count \\ 
        \cmidrule{1-2} \cmidrule{4-5} 
        VGG11 & 185 & ~ & Inception3 & 241 \\
        VGG16 & 176 & ~ & Inception4 & 243 \\
        VGG19 & 199 & ~ & & \\
        \bottomrule
    \end{tabular}
    \label{tab:vision-arch-1}
\end{table}
\vspace{-0.5em}
\begin{table}[h]
    \centering
    \caption{Resnet and U-Net architectures used for vision models}
    \begin{tabular}{lcclc}
    \toprule
    ResNet Models & Job Count & ~ & U-Net Models & Job Count \\ 
        \cmidrule{1-2} \cmidrule{4-5} 
        ResNet50 & 111 & ~ & U3-32 & 165\\
        ResNet50\_v1.5 & 91 & ~ & U3-64 & 159\\
        ResNet101 & 77 & ~ & U3-128 & 165\\
        ResNet101\_v2 & 54 & ~ & U4-32 & 163\\
        ResNet152 & 76 & ~ & U4-64 & 158\\
        ResNet152\_v2 & 54 & ~ & U4-128 & 157\\
        & & ~ & U5-32 & 158\\
        & & ~ & U5-64 & 158\\
        & & ~ & U5-128 & 148\\
        \bottomrule
    \end{tabular}
    \label{tab:vision-arch-2}
\end{table}
\vspace{-0.5em}
\begin{table}[h]
    \centering
    \caption{Architectures used for Natural Language Processing (NLP) and Graph Neural Networks (GNN)}
    \begin{tabular}{lcclc}
    \toprule
        NLP Model & Job Count & & GNN Model & Job Count \\ 
        \cmidrule{1-2} \cmidrule{4-5}
        Bert & 185 & & Dimenet & 33 \\
        DistillBert & 241 & & Schnet & 39 \\
        & & & PNA & 27 \\
        & & & NNConv & 32 \\
        \bottomrule
    \end{tabular}
    \label{tab:nlp-gnn-arch}
\end{table}

\end{document}